\documentclass[%
 reprint,
 amsmath,amssymb,
 aps,
pra,
]{revtex4-2}

\usepackage{graphicx}
\usepackage{dcolumn}
\usepackage{bm}
\usepackage{hyperref}

\usepackage{epsfig}
\usepackage{color}
\usepackage{bbm}
\usepackage{longtable}

\def\rmem#1#2#3{  \left\langle #1 \left\vert \left\vert  #2
                  \right\vert \right\vert #3 \right\rangle   }
\newcommand{\minus}[1]{ \mbox{$ (-1)^{#1} $}}

\newcommand{\CGC}[6]{ \mbox{$ \left( #1 #2 , #3 #4\,|\, #5 #6 \right) $} }

\newcommand{\SechsJ}[6]{ \mbox{$ %
            \arraycolsep0.25ex %
            \left\{ \begin{array}{ccc} %
                       #1 & #2 & #3 \vspace{0.5ex}\\%
                       #4 & #5 & #6 %
                   \end{array} \right\} $} }

\newcommand{\NeunJ}[9]{ \mbox{$ %
            \arraycolsep0.25ex %
            \left\{ \begin{array}{ccc} %
                       #1 & #2 & #3 \vspace{0.5ex}\\%
                       #4 & #5 & #6 \vspace{0.5ex}\\%
                       #7 & #8 & #9 %
                   \end{array} \right\} $} }

\begin{document}

\preprint{APS/123-QED}

\title{Vector parameters in atomic ionization by twisted light: polarization of electron and residual ion}

\author{Maksim D. Kiselev}
\affiliation{ 
Skobeltsyn Institute of Nuclear Physics, Lomonosov Moscow State University, 119991 Moscow, Russia;
}
 \affiliation{ 
 Faculty of Physics, Lomonosov Moscow State University, 119991 Moscow, Russia;
}
\affiliation{ 
 School of Physics and Engineering, ITMO University, 197101 Saint Petersburg, Russia;
}
 \affiliation{ 
Laboratory for Modeling of Quantum Processes, Pacific National University, 680035 Khabarovsk, Russia
}
\author{Elena V. Gryzlova}
\affiliation{ 
Skobeltsyn Institute of Nuclear Physics, Lomonosov Moscow State University, 119991 Moscow, Russia
}
\author{Alexei N. Grum-Grzhimailo}%
 \affiliation{ 
Skobeltsyn Institute of Nuclear Physics, Lomonosov Moscow State University, 119991 Moscow, Russia;
}
\affiliation{ 
 School of Physics and Engineering, ITMO University, 197101 Saint Petersburg, Russia
}

\date{\today}

\begin{abstract}
The electron and ion properties observed in a photo\-ionization inherit a symmetry properties of both a target and a radiation. 
Introducing a symmetry breaking in a photo\-ionization process one can expect to observe a noticeable variation of the vector correlation parameters of either outgoing photo\-electron or a residual ion. One of the ways to violate symmetry is to irradiate a matter by the twisted radiation which involves an additional screw.

In the paper we present the extension of the approach developed in~\cite{Kiselev2023} for the photo\-electron angular distribution to the other vector correlation parameters, exactly photo\-electron spin polarization, orientation and alignment of the residual ion. Usually two conditions are needed to produce polarized photo\-electrons: a system possesses a helix and a noticeable spin-orbital interaction. In the paper we investigate if a twisted light brings an additional helicity to a system. As an illustrative example we consider ionization of valence $4p$-shell of atomic krypton by circularly and linearly polarized Bessel light. The photo\-electron spin components are analyzed as a function of the cone angle of the twisted radiation.
\end{abstract}

\maketitle


\section{Introduction}
\label{Intro}

The technique of the laser radiation generation has been developing in the very diverse directions to reach higher frequencies, brighter intensities, shorter durations, pulse with a chirp or definite carrier envelope phase (CEP). Among of them the important role plays twisted light, i.e. pulses with definite projection of orbital angular momentum possessing  a nonuniform  intensity profile structure, a complex  surface of the constant phase and internal flow patterns~\cite{Mollina-Terriza2007,Bekshaev_2011}. Nowadays a twisted light became available in the broad range of energies from the optical region up to vacuum ultraviolet (VUV) range~\cite{Peele:02,Sasaki2008,Hemsing2011,Serbo2011,Serbo2011a,He:13, Bahrdt2013, Shen_2013,Ribic2014,hernandez2017,Bahrdt2013}. The manifold of generation techniques were developed and successfully applied: spiral phase plates~\cite{Sueda:04, BEIJERSBERGEN1994321}, holograms~\cite{Heckenberg:92},  $q$-plates~\cite{Karimi2009}, axicons~\cite{ARLT2000297}, integrated ring resonators~\cite{Cai2012}, on-chip devices~\cite{Yang2021}, Archimedean spirals~\cite{Kerber2018,Albar2023}.
Among a variety of twisted light types usually distinguish Laguerre–Gaussian~\cite{ALLEN1992,ALLEN1999} and  Bessel~\cite{Durnin:87,DurninPRL} beams. 

For the interaction of a twisted light with matter in the gas phase both experimental \cite{Babiker_2019} and theoretical researches have been performed for a different targets: atoms~\cite{Surzhykov2016,Kosheleva2020,Ramakrishna2022}, molecules~\cite{Araoka2005,Peshkov2015}, ions~\cite{MaH13,Seipt2016}.

The availability of UV twisted radiation makes possible the traditional photo\-electron spectroscopy, such as the measurements of photo\-electron angular distribution (PAD) or spin polarization with a twisted radiation. Because the twisting violates the selection rules~\cite{Picon:10} it may affect photo\-electron angular distribution and spin polarization in a very sophisticated way and the creation of general theory of photo\-ionization is highly desirable. In work~\cite{MaH13} the general formalism of ions photo\-ionization by the twisted Bessel light was developed for the hydrogen-like system with the Coulomb wave-functions.  The photo\-excitation of atoms by the Bessel beams have been already discussed in~\cite{Schulz2019}. Recently the  manifestation of the non-dipole effects in PADs due to irradiation of atoms by the Bessel light was analyzed in~\cite{Kiselev2023} and the present work is the extension of the approach to other vector correlation parameters of photo\-ionization.

It is reasonable to consider the vector characteristics of a system, such as angular distribution of the reaction products, in terms of certain spherical harmonic (or Wigner D-function)~\cite{Cooper1990}. Twisted radiation modifies the characteristics causing redistribution of terms with different ranks. Involving to consideration a particle polarization -- either photo\-electron or photo\-ion -- makes an observable physical picture much more vivid because of a competition between odd and even parameters. 

Being a fundamental property of particles spin carries the basis for many processes at different levels of matter: from the bottom with single elementary particles towards the top with the macroscopic objects. The possibilities of practical applications are quite wide: spintronics, ionic traps, laser cooling, quantum computers~\cite{Wolf2001,Rougemaille10,Johnson97,Heinzmann12,Lv19,Meng2023} and will most likely expand. 

Although gaseous targets are the most common species for photo\-ionization experiments, production and detection of spin-polarized electrons emitted in such processes carries difficulties in comparison with the ionization
of condensed matter since atoms and molecules in a gas target have predominantly random orientation and the medium itself is sparse.  

In the conventional photo\-ionization usually distinguish two necessary conditions for observation of nonzero spin polarization of photo\-electron:
(a) the process contains a helix (axial vector) and (b)
essential spin--orbit interaction. While the directions of the axial vectors correspond to possible photo\-electron spin components, a noticeable spin--orbit interaction makes it possible to resolve spin states of the residual ion or of the electron in continuum.

There are some well-known examples:\\
(i) The Fano effect for the spin orientation of the photoelectron ejected by circularly
polarized plane-wave light~\cite{Fano1969,Lubell1969,Kessler1970}. 
Electrons are to be collected over the full solid angle of 4$\pi$. The helix is provided by the
light helicity, and the only possible photoelectron spin component is parallel to the light beam propagation direction. On the contrary, the linearly
polarized plane-wave light does not generate the integral spin polarization at all because of lacking the helix.\\
(ii) Spin polarization observed in angle-resolved experiments on electron emission
from unpolarized atoms by the influence of linearly polarized plane-wave light (so-called \textit{dynamic
polarization})~\cite{Cherepkov1983}. Considering the electric dipole approximation,
the helix is provided by the vector product of the electric field and the direction of
the electron emission, therefore the only allowed spin component is normal to them~\cite{Cherepkov1979,Cherepkov1978}. \\
(iii) Same as (ii) but with circularly polarized plane-wave light (so-called \textit{polarization transfer}). 
In this case an additional helix appears due to the presence of light helicity and generally speaking all three possible photo\-electron spin components could be observed.

It leads to the idea that introduction of new distinguishable directions into a photo\-ionization process may violate the symmetry of a system  and bring new (or change already existing) photo\-electron spin components. This direction can be formed by a molecular frame~\cite{Cherepkov1983}, by a strong field~\cite{Barth13,Milocevic2016,Milocevic2018,Liu2018,Kabachnik2022} or by light consisting of components with proportional frequencies~\cite{Gryzlova20,Popova2021,Popova2022}.

In this work we expected to find spectacular manifestation of photo\-electron spin polarization during photo\-ionization of atom by twisted (Bessel) light. In our case the necessary helix is provided by the ``twistedness'' of light itself. Besides, we have chosen ionization of krypton $4p$-shell since ionic states $[\mathrm{Ar}]4s^24p^{-1}_{3/2}$ and $[\mathrm{Ar}]4s^24p^{-1}_{1/2}$ are energetically splitted quite enough ($\sim$0.67 eV,~\cite{Svens1988}) to be experimentally resolved.

The article is organized as follows. In Sec.~\ref{Section2} we briefly recall matrix element evaluation procedure in case of many-electron atom irradiation by twisted light and construct expressions for photo\-electron and photo\-ion statistical tensors in different coordinate systems. In Sec.~\ref{Section3} we present and discuss results on PADs and spin polarization of photoelectrons ejected by the circularly and linearly polarized Bessel light in $4p$-shell ionization of atomic krypton. The atomic units are used throughout the paper until otherwise specified.

\section{The general equations}
\label{Section2}

Let us consider an emission of a photo\-electron with momentum ${\bm p}$ and projection of spin  onto its propagation direction $m_s$:
\begin{equation}
\hbar\omega + A(\alpha_i J_i M_i) \rightarrow A^+(\alpha_f J_f M_f) + e^-({\bm p} m_s),
\label{eq:process}
\end{equation}
where $A$($A^+$) denotes  a state before (after) ionization i.e. an atom or an ion,  $\hbar\omega$ is the photon energy, $J_{i,f}$ and $M_{i,f}$ are the total angular momenta and their projections of the initial and final (atomic and ionic) states, and $\alpha_{i,f}$ are all other quantum numbers needed for the state specification.

\subsection{The twisted-wave matrix element}

We chose for analysis the photo\-ionization by the Bessel light propagating along the (quantization) $z$~axis. For this case, the Bessel state is characterized by the projections of the linear momentum $k_z$ and the total angular momentum (TAM) onto the $z$~axis $m_{\mathsf{tam}}$. The absolute value of the transverse momentum, $\kappa_{\perp} = \left| {\bf k}_{\perp} \right|$, is fixed; together with $k_z$ it defines the energy of the photons $\omega = c \sqrt{\kappa_{\perp}^2 + k_z^2}$. As shown in~\cite{MaH13}, this Bessel state is described by the vector potential 
\begin{equation} 
    \label{eq:18}
    {\bf A}_{\kappa_{\perp} m_{\mathsf{tam}} \lambda}^{\mathrm{tw}} = \int {\bf u}_{\lambda} e^{i{\bf kr}} 
    a_{\kappa_{\perp} m_{\mathsf{tam}}}({\bf k}_{\perp}) \frac{d^2 {\bf k}_{\perp}}{4 \pi^2} \, ,
\end{equation}
where
\begin{equation} 
    \label{eq:18a}
    a_{\kappa_{\perp} m_{\mathsf{tam}}}({\bf k}_{\perp}) =  (-i)^{m_{\mathsf{tam}}} e^{i m_{\mathsf{tam}} \phi_k} \sqrt{\frac{2 \pi}{k_{\perp}}}
    \delta(k_{\perp} - \kappa_{\perp}) \, ,
\end{equation}
and ${\bm u}_{\lambda}$ is the polarization vector with helicity $\lambda = \pm 1$.

The vector potential of twisted light  of any polarization is a combination of the basis components~(\ref{eq:18}) taken with appropriate weights $\epsilon_{\lambda}$: $\sum_{\lambda} \epsilon_{\lambda} {\bf A}_{\kappa_{\perp} m_{\mathsf{tam}} \lambda}^{\mathrm{tw}}$. For example  ``linearly polarized'' in $xz$-plane Bessel light can be obtained with $\epsilon_{\pm1}=\pm i/\sqrt{2}$~\cite{Schulz2020}. It should be mentioned that the polarization structure of the twisted beam is tricky question itself and was a subject of fruitful discussion~\cite{Quinteiro2015,Quinteiro2019} 

\begin{figure}
\centering
\includegraphics[width=0.49\textwidth]{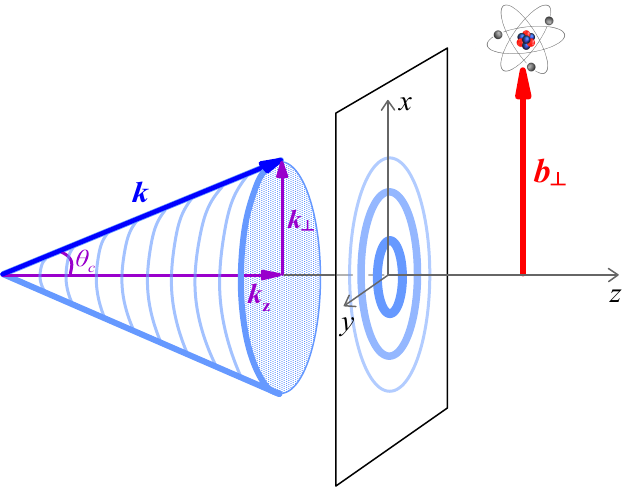}
\caption{The overview of the Bessel beam parameters, position of a target atom and schematic intensity profile in $xy$-plane. \label{fig0}}
\end{figure}

 The Bessel state characterized by eqs.~(\ref{eq:18})-(\ref{eq:18a}) can be understood as a coherent superposition of plane waves in momentum space with their wave vectors ${\bm k} = \left({\bm k}_\perp, k_z\right)$ lying on the surface of a cone with opening angle $\tan\theta_c = k_\perp/k_z$ (see Fig.~\ref{fig0}).

Using the vector potential (\ref{eq:18}), we obtain  the matrix element for photo\-ionization:  
\begin{eqnarray}
    \label{eq:matrix_element_twisted}
    &&M^{(\rm tw)}_{M_i \lambda \, m_{\mathsf{tam}} M_f}\left({\bm p}; \theta_c, \, {\bm b} \right) \nonumber\\
  &=&\int \, a_{\kappa_{\perp} m_{\mathsf{tam}}}({\bf k}_{\perp}) \, {\rm e}^{-i {\bm k}_\perp {\bm b}_{\perp}} \, M^{(\rm pl)}_{M_i \lambda M_f}\left({\bm k}, {\bm p} \right)\frac{d^2 {\bf k}_{\perp}}{4 \pi^2}  \,  \, ,
\end{eqnarray}
where $M^{(\rm pl)}_{M_i \lambda M_f}\left({\bm k}, {\bm p} \right)$ is the conventional plane-wave matrix element in $jj$-coupling scheme. Therefore:

\begin{eqnarray}
 & &  \label{eq:matrix_element_plane_wave_final_2}
     M^{(\rm pl)}_{M_i \lambda M_f}\left({\bm k}, {\bm p} \right) = \sqrt{2\pi} \, \sum\limits_{LMp}  \, \sum\limits_{\kappa \mu J_t M_t} i^{L} \, \left(i \lambda \right)^{p} \nonumber\\
    && \frac{[lL] }{[J_t]} \,\CGC{l}{0}{\frac{1}{2}}{m_s}{j}{m_s} \,\CGC{J_f}{M_f}{j}{\mu}{J_t}{M_t}\times \nonumber\\
  &&    \CGC{J_i}{M_i}{L}{M}{J_t}{M_t}  \, D^{j *}_{\mu m_s}(\hat{\bm p}) \, D^L_{M \lambda}(\hat{{\bm k}}) \nonumber\\
     &&\rmem{(\alpha_f J_f, \epsilon \kappa) J_t}{H_{\gamma}(pL)}{\alpha_i J_i} \, ,
\end{eqnarray}
and $D^{j}_{m m'}(\hat{{\bm k}})$ is the Wigner D-function (see, for example,~\cite{Balashov2000}); $\hat{{\bm k}} = (\phi_k, \theta_k, 0)$ defines the direction of the incident (plane-wave) photon; $\hat{\bm p} = (\phi_p, \theta_p, 0)$ is the propagation direction of photoelectron; $\kappa = j+1/2$ for $l = j \pm 1/2$ ($l$ is the orbital angular momentum of the electron) defines the Dirac angular-momentum quantum number. Note that projection of total photo\-electron momentum $\mu$ and the whole system momentum $M_t$ are in the laboratory coordinate system.  The notation $[abc...] \equiv \sqrt{(2a+1)(2b+1)(2c+1)...}$\, and standard designation for the Clebsch-Gordan coefficients are used. Operator $H_{\gamma}(pL)$ is responsible for the interaction between atomic electron and magnetic ($p=0$) or electric ($p=1$) photon with multipolarity $L$. Note that the reduced matrix element $\rmem{(\alpha_f J_f, \epsilon \kappa) J_t}{H_{\gamma}(pL)}{\alpha_i J_i}$ includes the scattering phase dependence (see~\cite{Kiselev2023} for the details). In Eq.~(\ref{eq:matrix_element_twisted}) the factor ${\rm e}^{-i {\bm k}_\perp {\bm b}_{\perp}}$ with ${\bm b}_\perp = \left(b_x, b_y \right)$ specifies the position of the target atom regarding to the quantization axis of the incident light (see Fig.~\ref{fig0}).

Applying two vector potentials (\ref{eq:18}), obtained for different TAM projections and different helicities one can derive the matrix element of photo\-ionization by twisted light of any polarization:
\begin{eqnarray}
    \label{eq:matrix_element_photoionization_linear}
    M^{(\rm tw)}_{M_i M_f }\left({\bm p}; \theta_c, \, {\bm b} \right)&=& \\
   &=& \sum_{\lambda}  \epsilon_{\lambda} \,  M^{(\rm tw)}_{M_i \lambda \, m_{\mathsf{tam}} = m +\lambda \, M_f}\left({\bm p}; \theta_c, \, {\bm b} \right)  \, ,\nonumber
\end{eqnarray}
in terms of the matrix elements (\ref{eq:matrix_element_twisted}).

\subsection{Observable parameters}

We consider the target (atom) being initially unpolarized and distinguish cases when  the polarization of photo\-electron and photo\-ion is detected. Therefore we should
average the photo\-emission probability over initial magnetic quantum number $M_i$ and sum incoherently either over
final magnetic quantum number $M_f$ or $m_s$. Evaluation of the angle resolved spin polarization of a macroscopic (i.e. consists of atoms randomly and uniformly distributed within the $xy$-plane) atomic target can be performed by averaging the product of matrix elements corresponding to specific quantum numbers of electron spin projection $m_s,m_s'$ (\ref{eq:matrix_element_twisted}) over the impact parameter:
\begin{widetext}
\begin{eqnarray}
    \label{eq:crpss_section_linear}
    \frac{{\rm d}\sigma_{m_sm'_s}^{\rm (tw)}}{{\rm d}\Omega_p}\left(\theta_p, \phi_p ; \, \theta_c\right) &=& \mathcal{N} \frac{1}{2J_i + 1} \, \sum\limits_{M_i M_f} \sum\limits_{\lambda, \lambda'} \epsilon_{\lambda}\epsilon^{\ast}_{\lambda'} 
    \int M^{(\rm tw)}_{M_i \lambda \, m_{\mathsf{tam}} = m +\lambda \, M_f}\left({\bm p}; \theta_c, \, {\bm b} \right) \, M^{(\rm tw)\ast}_{M_i \lambda' \, m_{\mathsf{tam}} = m +\lambda' \, M_f}\left({\bm p}; \theta_c, \, {\bm b} \right) \, \frac{{\rm d}{\bm b}_{\perp}}{\pi R^2} \nonumber \\
 &   =& \mathcal{N} \frac{1}{2J_i + 1}  \, \sum\limits_{M_i M_f} 
    \sum\limits_{\lambda, \lambda'} \epsilon_{\lambda}\epsilon^{\ast}_{\lambda'} \int M^{(\rm pl)}_{M_i \, \lambda \, M_f}({\bm k}, {\bm p}) \, M^{(\rm pl) *}_{M_i \, \lambda' \, M_f}({\bm k}, {\bm p}) \, {\rm e}^{i (\lambda - \lambda') \varphi_k}  \, \frac{d {\varphi_k}}{2 \pi}  \, ,
\end{eqnarray}
where the parameter $R$ defines the ``size'' of a target and is assumed to be much larger than the characteristic size of the Bessel beam intensity profile patterns. The evaluation of the prefactor $\mathcal{N}$ requires re--definition of the concept of cross section~\cite{ScF14} and is not a subject of the current investigation, because here we interested in the non-dimensional parameters.

To simplify Eq.~(\ref{eq:crpss_section_linear}) we first substitute the matrix elements (\ref{eq:matrix_element_plane_wave_final_2}) and obtain  integral

\begin{equation} \label{eq:igr}
\int \!d \phi_k \, e^{i(\lambda-\lambda') \phi_k} D^{s}_{\lambda-\lambda',\, q}(\hat{\bm k}^{-1}) = 2 \pi \, d^s_{q q}(\theta_c) 
\delta_{q,\, \lambda-\lambda'}\, .
\end{equation}
 One can see that parameter ${\bm b}_{\perp}$ which may be essential to account a complex internal structure of a twisted light consisting of concentric rings of high and low intensity is smeared out for a macroscopic target.
 
In such a way we obtain expression for the component of the angle-resolved spin density matrix elements:

\begin{eqnarray}
 \frac{{\rm d}\sigma^{\rm (tw)}_{m_s,m_s'}}{{\rm d}\Omega_p}\left(\theta_p, \phi_p ; \, \theta_c\right)&=&\frac{2 \pi \mathcal{N}}{2J_i + 1} 
\sum_{k,q,k_l}  D^{k\ast}_{q\, q_s}(\phi_p,\theta_p,0) \, d^k_{qq}(\theta_c) \sum_{LL'pp'} \minus{1/2-m'_s} \delta_{q, \xi} \rho_{k\xi}[pL,p'L']
\nonumber\\
&\times&\CGC{1/2}{\,m_s}{1/2}{-m'_s}{k_s}{q_s}\CGC{k_l}{0}{k_s}{q_s}{k}{q_s}B^{Lp,L'p'}[k_l,k_s,k],
\label{eq:spin_polar_component}
\end{eqnarray}
where the dynamical parameter
\begin{eqnarray} 
 B^{Lp,L'p'}[k_l,k_s,k]& = & \sum_{\kappa \kappa' J_t J'_t}
\minus{J_i-J_f+L'-j'+l'}  [ll'jj'LL'J_t J'_t k_l k_s] \CGC{l}{0}{l'}{0}{k_l}{0} \NeunJ{l}{1/2}{j}{l'}{1/2}{j'}{k_l}{k_s}{k} \nonumber \\
&\times& 
\SechsJ{J_t}{J'_t}{k}{L'}{L}{J_i} \SechsJ{J_t}{J'_t}{k}{j'}{j}{J_f}
 \rmem{(\alpha_f J_f, \epsilon \kappa) J_t}{H_{\gamma}(pL)}{\alpha_i J_i} \,
\rmem{(\alpha_f J_f, \epsilon \kappa') J'_t}{H_{\gamma}(p'L')}{\alpha_i J_i}^{\ast}
\label{eq:Bparameter_general}
\end{eqnarray}
does not depend on the polarization parameters and
\begin{eqnarray}
\rho_{k\xi}[pL,p'L'] &=& \sum_{\lambda\lambda' } \epsilon_{\lambda} \epsilon^{\ast}_{\lambda'}  i^{L-L'}(i \lambda)^p (-i \lambda')^{p'} \minus{L'+\lambda'} \CGC{L}{\lambda}{L'}{-\!\!\lambda'}{k}{\xi}
\label{eq:rho_general}
\end{eqnarray}
is the statistical tensor of photon with definite multipolarity and type.
Therefore the photo\-electron statistical tensor as a function of $\left(\theta_p, \phi_p ; \, \theta_c\right)$:

\begin{eqnarray} \label{eq:Stat}
 \rho^{\rm (tw)}_{k_sq_s} [1/2, 1/2]& = &
\sum_{kq} D^{k\ast}_{q\, q_s}(\phi_p,\theta_p,0) \, d^k_{qq}(\theta_c) \sum_{LL'pp'} 
\rho_{kq}[pL,p'L']\CGC{k_l}{0}{k_s}{q_s}{k}{q_s}B^{Lp,L'p'}[k_l,k_s,k]\,.
\end{eqnarray}

For constructions~(\ref{eq:rho_general}) the usual tensor's  permutation rule is fair: $\rho_{k\xi}[pL,p'L']=(-1)^{L-L'+\xi}\rho^{\ast}_{k-\xi}[p'L', pL]$. Additionally for the dynamical parameters~(\ref{eq:Bparameter_general}): $B^{Lp, L'p'}[k_l,k_s,k]=(-1)^{k_l+k_s+k+L-L'}B^{L'p', Lp}[k_l,k_s,k]^{\ast}$.  
The permutation connection for~(\ref{eq:Stat}): $\rho_{k_sq_s}^{\rm (tw)}[1/2, 1/2]=(-1)^{q_s}\rho_{k_s-q_s}^{{\rm (tw)}\ast}[1/2, 1/2]$. If one puts $k_s=0\,,q_s=0$ into~(\ref{eq:Stat}), the equation turns to the PAD and coincides with equation (19) or (27) from~\cite{Kiselev2023} up to normalization. Since we interested in dimensionless parameters of angular anisotropy and spin polarization, we do not normalize density matrix and its trace is proportional to ionization probability. The form of eqs.~(\ref{eq:spin_polar_component}) and (\ref{eq:Stat}) is convenient because clearly separates the dynamical factor (\ref{eq:Bparameter_general}) depending on target details (atom, ionizing shell, photon energy, etc.) from the kinematic (geometrical) factor (\ref{eq:rho_general}) depending on polarization, orientation and others.

Following the similar way with minor difference in summing over $m_s$ instead of $M_f$ and integrating over the electron emission angle one may easy obtain the statistical tensor components for the residual ion:
\begin{eqnarray} \label{eq:st_ion}
 \rho^{\rm (tw)}_{k_fq_f} [J_f, J_f]& = &
\delta_{kk_f}\delta_{qq_f}  d^k_{qq}(\theta_c) \sum_{LL'pp'} 
\rho_{kq}[pL,p'L']\bar{B}^{Lp,L'p'}[k] \nonumber\\
\end{eqnarray}
with the dynamical parameter:

\begin{eqnarray} 
\bar{ B}^{Lp,L'p'}[k]& = & \sum_{\kappa \kappa' J_t J'_t}
\minus{J_i+J+L'-J_f-j-J'}  [LL'J_t J'_t]  \nonumber \\
&\times& 
\SechsJ{J_t}{J'_t}{k}{L'}{L}{J_i} \SechsJ{J_t}{J'_t}{k}{J_f}{J_f}{j}
 \rmem{(\alpha_f J_f, \epsilon \kappa) J_t}{H_{\gamma}(pL)}{\alpha_i J_i} \,
\rmem{(\alpha_f J_f, \epsilon \kappa') J'_t}{H_{\gamma}(p'L')}{\alpha_i J_i}^{\ast}.
\label{eq:Bparameter_ion}
\end{eqnarray}
\end{widetext}
The ratios of the components (\ref{eq:st_ion}) define polarization (orientation and alignment) of the ion.

The Cartesian components of electron spin polarization are related to the statistical tensors in the same coordinate system:
\begin{eqnarray}
S_z&=&\frac{\rho_{10}[1/2, 1/2]}{\rho_{00}[1/2, 1/2]}\,\label{eq:sz} \,, \\
S_x&=&-i\,\frac{\rho_{11}[1/2, 1/2]+\rho_{1-\hspace{-1pt}1}[1/2, 1/2]}{\sqrt{2}\rho_{00}[1/2, 1/2]}\,\label{eq:sy} \,, \\
S_y&=&-\,\frac{\rho_{11}[1/2, 1/2]-\rho_{1-\hspace{-1pt}1}[1/2, 1/2]}{\sqrt{2}\rho_{00}[1/2, 1/2]}\,\label{eq:sx} \,,
\end{eqnarray}

and PAD is defined by:
\begin{eqnarray}
\label{eq:PADstattens}
W &=& \sqrt{2}\rho_{00}[1/2, 1/2]\,.
\end{eqnarray}

\begin{figure}
\centering
\includegraphics[width=0.4\textwidth]{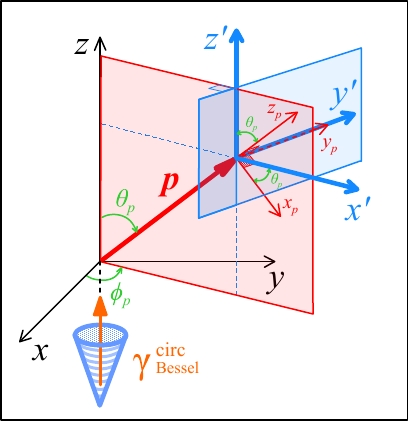}
\caption{Coordinate systems used in analysis of ionization by circularly polarized Bessel beam: laboratory coordinate system $S(x,y,z)$; photo\-electron coordinate system $S_p(x_p,y_p,z_p)$ in which statistical tensors (\ref{eq:Circ_PAD})-(\ref{eq:Circ_Sxy}) are written; coordinate system $S'(x',y',z')$ obtained from $S_p$ after the Euler's angles rotation $\omega_{\rm{circ}}=\{0;\,-\theta_p;\,0\}$ in which statistical tensors (\ref{eq:rho00rot_circ})-(\ref{eq:rho11rot_circ}) are written.
\label{fig:CScirc}}
\end{figure}

\begin{figure}
\centering
\includegraphics[width=0.45\textwidth]{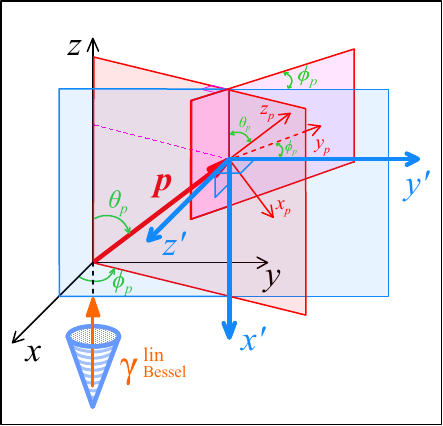}
\caption{Coordinate systems used in analysis of ionization by linearly polarized Bessel beam: laboratory coordinate system $S(x,y,z)$; photoelectron coordinate system $S_p(x_p,y_p,z_p)$ in which statistical tensors (\ref{eq:Lin_PAD})-(\ref{eq:Lin_spin}) are written; coordinate system $S'(x',y',z')$ obtained from $S_p$ after two subsequent Euler's angles rotations: $\omega_{\rm{lin}}^{(1)}=\{0;\,-\theta_p;\,-\phi_p\}$ and $\omega_{\rm{lin}}^{(2)}=\{0;\,\pi/2;\,0\}$ in which statistical tensors (\ref{eq:rho00rot_lin})-(\ref{eq:rho11rot_lin}) are written.
\label{fig:CSlin}}
\end{figure}

For the coordinate system used in (\ref{eq:Stat}) the $z$-component is along the photo\-electron emission direction, the $x$-component is in the plane formed by the direction of electro\-magnetic field propagation and of electron emission ({\it tangential} component), the $y$-component is orthogonal to these vectors ({\it normal} component) (see coordinate systems marked by red in Figs.~\ref{fig:CScirc} and~\ref{fig:CSlin}).

In dipole approximation the explicit forms of eqs.~(\ref{eq:Stat}) are:
\begin{eqnarray}
\rho_{00}^{\rm (tw)}&=&\rho_{00}^{E1}B[0,0,0]+\nonumber\\
&+&D^{2\ast}_{q0}(\phi_p,\theta_p,0) \, d^2_{qq}(\theta_c) \rho_{2q}^{E1}B[2,0,2]   \\
\rho_{10}^{\rm (tw)}&=&D^{1\ast}_{q0}(\phi_p,\theta_p,0) \, d^1_{qq}(\theta_c)\rho_{1q}^{E1}B[0,1,1]+\nonumber\\
&-&\sqrt{\frac{2}{5}}D^{1\ast}_{q0}(\phi_p,\theta_p,0) \, d^1_{qq}(\theta_c)\rho_{1q}^{E1}B[2,1,1]   \\
\rho_{11}^{\rm (tw)}&=&D^{1\ast}_{q1}(\phi_p,\theta_p,0) \, d^1_{qq}(\theta_c)\rho_{1q}^{E1}B[0,1,1]+\nonumber\\
&+&\frac{1}{\sqrt{10}}D^{1\ast}_{q1}(\phi_p,\theta_p,0) \, d^1_{qq}(\theta_c)\rho_{1q}^{E1}B[2,1,1]+\nonumber\\
&-&\frac{1}{\sqrt{2}}D^{2\ast}_{q1}(\phi_p,\theta_p,0) \, d^2_{qq}(\theta_c)\rho_{2q}^{E1}B[2,1,2]
\end{eqnarray}
Here and below for the brevity we put $B^{E1,E1}[k_l,k_s,k]\equiv B[k_l,k_s,k]$, $\rho_{kq}^{\rm (tw)}[1/2,1/2] \equiv \rho_{kq}^{\rm (tw)}$ and $\rho_{kq}[E1,E1]\equiv\rho_{kq}^{E1}$. According to the permutation rules for dynamical parameters it is obvious that $B[0,0,0]$, $B[2,0,2]$, $B[0,1,1]$ and $B[2,1,1]$ have only real parts, but $B[2,1,2]$ is completely imaginary. 

For the circularly polarized beam $\epsilon_{\pm 1}=1$, $\rho_{00}^{E1}=1/\sqrt{3}$, $\rho_{10}^{E1}=\pm 1/\sqrt{2}$, $\rho_{20}^{E1}=1/\sqrt{6}$ and thus: 

\begin{eqnarray}
\rho_{00}^{\rm (tw)}&=&\frac{1}{\sqrt{3}}B[0,0,0]+
\frac{1}{\sqrt{6}}P_2(\cos\theta_p) \, P_2(\cos\theta_c) B[2,0,2] \nonumber \\  \label{eq:Circ_PAD} \\
\rho_{10}^{\rm (tw)}&=& \frac{\pm1}{\sqrt{2}} \, \cos\theta_p \, \cos\theta_c \left(B[0,1,1]
- \sqrt{\frac{2}{5}}B[2,1,1]\right) \label{eq:Circ_Sz}  \\
\rho_{11}^{\rm (tw)}&=&-\frac{1}{2} \, \sin\theta_p\left\{\mp \, \cos\theta_c \left( B[0,1,1]+\frac{1}{\sqrt{10}} \,  \, B[2,1,1]\right) \right.\nonumber\\
&+&\left. \frac{1}{\sqrt{2}}\cos\theta_p \, 
P_2(\cos\theta_c)B[2,1,2]\right\} \label{eq:Circ_Sxy}
\end{eqnarray}
Similar to the case of plane wave ionization eqs.~(\ref{eq:Circ_PAD})-(\ref{eq:Circ_Sxy}) possess the axial symmetry with respect to the beam propagation direction. No additional spin component appears, but the dependency of them on the opening angle $\theta_c$ is different. 
For example, the imaginary part of eq.~(\ref{eq:Circ_Sxy}) turns to zero at $\theta_c=\arccos(1/\sqrt{3})$, while the real part conserves and the electron is polarized in $xz$-plane.

The orientation and alignment of the residual ion for twisted and plane radiation are connected as:
\begin{eqnarray}
 \mathcal{A}^{\rm (tw)}_1&=\frac{ \rho^{\rm (tw)}_{10} [J_f, J_f]}{ \rho^{\rm (tw)}_{00} [J_f, J_f]} &=\cos\theta_c\,\mathcal{A}_1\,;\label{eq:ion_circ1}\\   
 \mathcal{A}^{\rm (tw)}_2&=\frac{ \rho^{\rm (tw)}_{20} [J_f, J_f]}{ \rho^{\rm (tw)}_{00} [J_f, J_f]} &=\left(\frac{3\cos^2\theta_c-1}{2}\right)\mathcal{A}_2\,.\label{eq:ion_circ2}
\end{eqnarray}
Therefore the orientation drops with cone angle slower than the alignment. Let us remind a presentation of polarization in terms of population $n_{M_f}$ of sub\-levels with definite magnetic quantum number $M_f$:
\begin{eqnarray} \label{eq:populat}
\mathcal{A}_{k0}&=&[J_f]\frac{\sum_{M_f} (-1)^{J_f-M_f}\CGC{J_f}{M_f}{J_f}{-M_f}{k}{0}n_{M_f}}{\sum_{M_f} n_{M_f}}\,. \nonumber\\
\end{eqnarray}
Therefore eqs.~(\ref{eq:ion_circ1})-(\ref{eq:ion_circ2}) may be understood as an averaged over a macroscopic target population of different magnetic sub\-levels.

Statistical tensors under the Euler angles' rotation $\omega=\{\alpha;\,\beta;\,\gamma\}$ transforms as~\cite{Balashov2000}:
\begin{equation}
\label{eq:rotation_stattensor}
\widetilde{\rho}_{k'q'}[j, j']=\delta_{kk'}\,\sum_{q}\rho_{kq}[j, j']D^{k*}_{qq'}(\omega)
\end{equation}

In order to discuss the case of photo\-ionization by circularly polarized Bessel beam in laboratory coordinate system it is sufficient to transform photoelectron coordinate system $S_p(x_p,y_p,z_p)$ into the system $S'(x',y',z')$ in such a way that $z'$ axis align with $z$ axis. That could be done by Euler's rotation $\omega_{\rm{circ}}=\{0;\,-\theta_p;\,0\}$ (see Figure~\ref{fig:CScirc}), and then statistical tensors' components in case of circular polarization become:
\begin{eqnarray}
\widetilde{\rho}_{00}^{\,\rm (tw)}&=&\rho_{00}^{\rm (tw)} \label{eq:rho00rot_circ} \\
\widetilde{\rho}_{10}^{\,\rm (tw)}&=&\sqrt{2}\sin\theta_p\cdot \mathrm{Re}\rho_{11}^{\rm (tw)}+\rho_{10}^{\rm (tw)}\cos\theta_p \nonumber \\
&=&\pm\frac{\cos\theta_c}{\sqrt{2}}\left( B[0,1,1]-\sqrt{\frac{2}{5}}B[2,1,1]P_2(\cos\theta_p) \right) \nonumber\\ \label{eq:rho10rot_circ} \\
\widetilde{\rho}_{11}^{\,\rm (tw)}&=&\cos\theta_p\cdot\mathrm{Re}\rho_{11}^{\rm (tw)}-\frac{\sin\theta_p}{\sqrt{2}}\rho_{10}^{\rm (tw)}+i\,\mathrm{Im}\rho_{11}^{\rm (tw)} \nonumber\\
&=&\pm\frac{\sin\theta_p\cos\theta_p}{2\sqrt{2}}\left(\frac{3}{\sqrt{5}}\cos\theta_cB[2,1,1] \right. \nonumber \\
 &\mp& \bigg. i\,P_2(\cos\theta_c)\mathrm{Im}B[2,1,2] \bigg) \label{eq:rho11rot_circ} 
\end{eqnarray}
The part of (\ref{eq:rho10rot_circ}) without angle dependency on $\theta_p$ is the only one component which conserves at integration over photo\-emission direction, which is very similar to the Fano effect  \cite{Fano1969,Lubell1969,Kessler1970} in case of ionization by the plane-wave beam. The second term of (\ref{eq:rho10rot_circ}) provides the angular dependency of the spin component oriented along the beam propagation direction. The form of (\ref{eq:rho11rot_circ}) allows for clear distinguishing the real part, responsible for spin component in the plane of electron emission, and the imaginary part, responsible for the component orthogonal to it. 

For the light linearly polarized along $x$-axis $\epsilon_{\pm1}=\pm i/\sqrt{2}$, $\rho_{00}^{E1}=1/\sqrt{3}$, $\rho_{20}^{E1}=1/\sqrt{6}$, $\rho_{2\pm 2}^{E1}=-1/2$ and thus:

\begin{eqnarray}
\label{eq:Lin_PAD} \rho_{00}^{\rm (tw)}&=&\frac{1}{\sqrt{3}}B[0,0,0]+\\
&+&\frac{1}{\sqrt{6}}B[2,0,2]\left\{\vphantom{\frac{1}{2}}1-3\sin^2\theta_p\cos^2\phi_p
\right.\nonumber\\
&-&6\sin^2(\theta_c/2)( \cos^2\theta_p -  \sin^2\theta_p\cos^2\phi_p)\nonumber\\
&+&\left. \frac{3}{2}\sin^4(\theta_c/2)(5\cos^2\theta_p-1-2\sin^2\theta_p \cos^2\phi_p)\right\} \nonumber \\
\label{eq:Lin_spin} \rho_{11}^{\rm (tw)}&=&-\frac{1}{\sqrt{2}}B[2,1,2]\sin\theta_p \\
&\cdot&\left\{\vphantom{\frac{1}{2}}\!\cos\theta_p\cos^2\phi_p+i\cos\phi_p\sin\phi_p\right.\nonumber\\
  &-&2\sin^2(\theta_c/2)(\cos\theta_p(1+\cos^2\phi_p)+i\cos\phi_p\sin\phi_p)\nonumber\\
  &+&\left.\sin^4(\theta_c/2)(\cos\theta_p \, (\frac{5}{2}+\cos^2\phi_p)+i\cos\phi_p\sin\phi_p)\right\} \nonumber
\end{eqnarray}

The ratio $-\sqrt{2}B[2,0,2]/B[0,0,0]$ at the $\theta_c=0$ is the conventional angular anisotropy parameter $\beta$ of PAD.  Three important notes should be mentioned: (A) there is no tensor component with $q_s=0$ which means that there is no spin component oriented along photo\-electron emission direction, as it is for plane-wave ionization \cite{Cherepkov1978,Kabachnik2004}; (B) accounting that $B[2,1,2]$ is imaginary, the coordinates $n'_x=\{\sin\theta_p\cos\phi_p,\sin\theta_p\sin\phi_p,\cos\theta_p\}$, $n'_y=\{-\sin\phi_p,\cos\phi_p,0\}$ and substituting $\theta_c=0$ into eq.~(\ref{eq:Lin_spin}) one can check that there is no spin component oriented along polarization vector in case of plane-wave ionization~\cite{Cherepkov1978,Kabachnik2004}; (C) since real and imaginary parts of  eq.~(\ref{eq:Lin_spin}) depend on the opening angle $\theta_c$ in a different way additional spin component oriented along linear polarization vector appears for the twisted light.

For the linearly polarized twisted light there are only second rank components of the polarization:
\begin{eqnarray}
\mathcal{A}^{\rm (tw)}_{20}&=&\left(\frac{3\cos^2\theta_c-1}{2}\right)\mathcal{A}_{20}\,;\label{eq:ion_lin1}\\
\mathcal{A}^{\rm (tw)}_{2\pm 2}&=&\cos^4(\theta_c/2)\,\mathcal{A}_{2\pm 2}\,.\label{eq:ion_lin2}
\end{eqnarray}
As it is for circularly polarized light the dependency of different component of polarization on the cone angle is different.

In order to look carefully to the component of photo\-electron spin polarization oriented along the polarization it is constructive to switch into the system where new axis $z'$ is along to polarization vector (original $x$-axis). The corresponding rotation is performed with usage eq.~(\ref{eq:rotation_stattensor}) via two subsequent rotations: $\omega_{\rm{lin}}^{(1)}=\{0;\,-\theta_p;\,-\phi_p\}$ and $\omega_{\rm{lin}}^{(2)}=\{0;\,\pi/2;\,0\}$ (see Figure~\ref{fig:CSlin}). After that transformation the statistical tensors' components become:

\begin{eqnarray}
\widetilde{\rho}_{00}^{\,\rm (tw)}&=&\rho_{00}^{\rm (tw)} \label{eq:rho00rot_lin} \\
\widetilde{\rho}_{10}^{\,\rm (tw)}&=&-\sqrt{2}\left(\cos\theta_p\cos\phi_p\cdot\mathrm{Re}\rho_{11}^{\rm (tw)}+\sin\phi_p\cdot\mathrm{Im}\rho_{11}^{\rm (tw)}\right) \nonumber \\
&=& \frac{\cos\theta_p\sin\theta_p\sin\phi_p}{2}\mathrm{Im}B[2,1,2]\sin^2\frac{\theta_c}{2} \nonumber\\
&&\cdot\left(5\sin^2\frac{\theta_c}{2}-4 \right)\label{eq:rho10rot_lin} \\
\widetilde{\rho}_{11}^{\,\rm (tw)}&=&\sin\theta_p\cdot\mathrm{Re}\rho_{11}^{\rm (tw)} \nonumber\\
&+&i\left(\cos\phi_p\cdot\mathrm{Im}\rho_{11}^{\rm (tw)}-\sin\phi_p\cos\theta_p\cdot\mathrm{Re}\rho_{11}^{\rm (tw)} \right) \nonumber\\
&=& \frac{\sin\theta_p\cos\phi_p}{\sqrt{2}}\mathrm{Im}B[2,1,2] \left\{ \sin\theta_p\sin\phi_p\cos^4\frac{\theta_c}{2} \right. \nonumber\\
&-& \left. i\cos\theta_p \left( 1-4\sin^2\frac{\theta_c}{2}+\frac{7}{2}\sin^4\frac{\theta_c}{2}\right) \right\} \label{eq:rho11rot_lin} 
\end{eqnarray}

All spin components arising for the linearly polarized twisted beam are determined by the only one dynamical parameter $B[2,1,2]$. The component (\ref{eq:rho10rot_lin}) arises only for the twisted radiation and does not exist for the plane one. 

While the physical meaning of a photo\-electron spin is unambiguous, the orientation and alignment need in additional discussion for the macroscopic targets. The ionic polarization parameters may be interpreted in the same way as for electron: assuming that a detector registers ions in the state with particular projection $M_f$ to a quantization axis and casting an appropriate  combination~(\ref{eq:populat}).
On the other hand in reality an ionic polarization is detected via some subsequent process. Our analysis has shown that if this subsequent process applying to determine the polarization does not involve the twisted light (for example, Auger decay or fluorescence) then the target averaged parameters (\ref{eq:ion_circ1}),~(\ref{eq:ion_circ2}),~(\ref{eq:ion_lin1}),~(\ref{eq:ion_lin2}) may be used in a usual way. However, if the process applying to determine polarization  involves the twisted light (for example, subsequent ionization of the ion or laser-assisted Auger decay) then the equation became inapplicable.

\section{Results and discussion}
\label{Section3}

\begin{figure}
\centering
\includegraphics[width=0.47\textwidth]{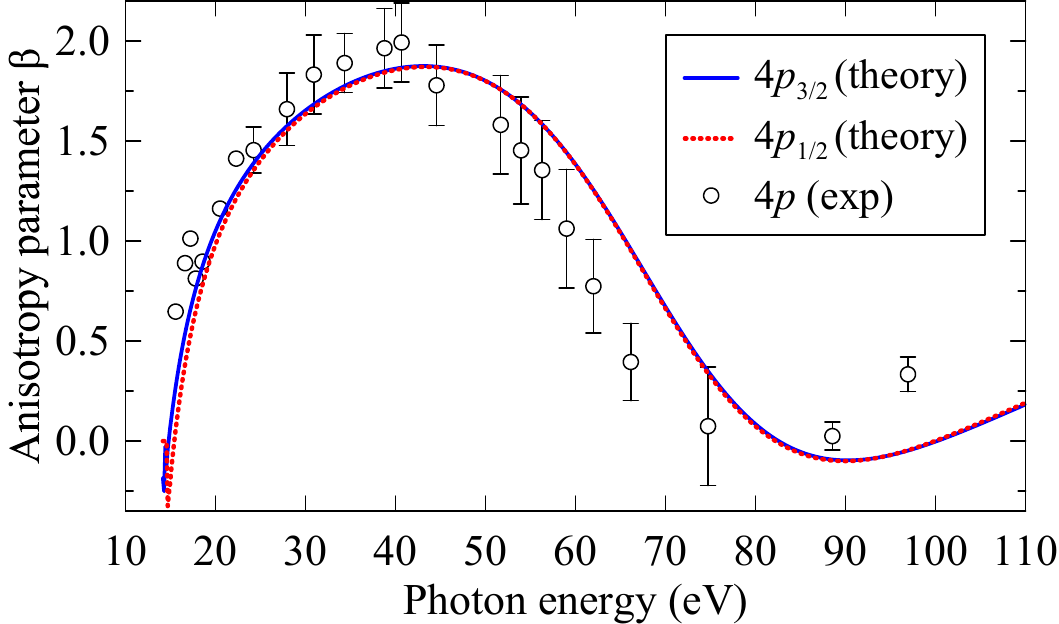}
\caption{Simulated according to present theoretical model (see text) anisotropy parameter $\beta$ for the PAD due to photoionization of krypton atom for final ionic states Kr$^+4p^5$ with $J_f=3/2$ (solid blue line) and $J_f=1/2$ (dotted red line). Experimental data are taken from~\cite{Miller1977}.
\label{fig:beta4p}}
\end{figure}

To illustrate the effects of twisting inprinted into a vector correlation, i.e. photo\-electron angular distribution and spin polarization we consider valence $4p$-shell ionization of krypton in the energy range below 90~eV not reaching the $3d$-shell ionization thresholds~\cite{Svens1988}.  Initial state of neutral krypton was prepared using the multi\-configuration Hartree-Fock method as a pure $[\mathrm{Ar}]4s^24p^6\,^1S_0$ state by means of MCHF code~\cite{Froese97}. After that $[\mathrm{Ar}]$ core was “frozen” and used to obtain $4s$ and $4p$ orbitals of final ionic state $[\mathrm{Ar}]4s^24p^5$ optimizing them on the $^2P$ term. The subsequent Breit-Pauli diagonalization procedure~\cite{Zatsarinny2013} was used in order to construct $j$-splitted states $[\mathrm{Ar}]4s^24p^5\,^2P_{3/2,\,1/2}$. After that we applied the B-spline R-matrix approach~\cite{Zatsarinny2006} to calculate photo\-ionization amplitudes of the process~(\ref{eq:process}) with the use of measured $4p_{3/2,1/2}$ subshell ionization thresholds~\cite{Svens1988}. The quality of the model is sufficient to describe reasonably the behavior of the dipole anisotropy parameter $\beta$ of the PAD due to krypton $4p$-shell photoionization (see Fig.~\ref{fig:beta4p}). The model predicts the dipole Cooper minimum at $\sim$82~eV which is close enough to more precise calculations of~\cite{Fritzsche2009}.

\begin{figure}
\centering
\includegraphics[width=0.5\textwidth]{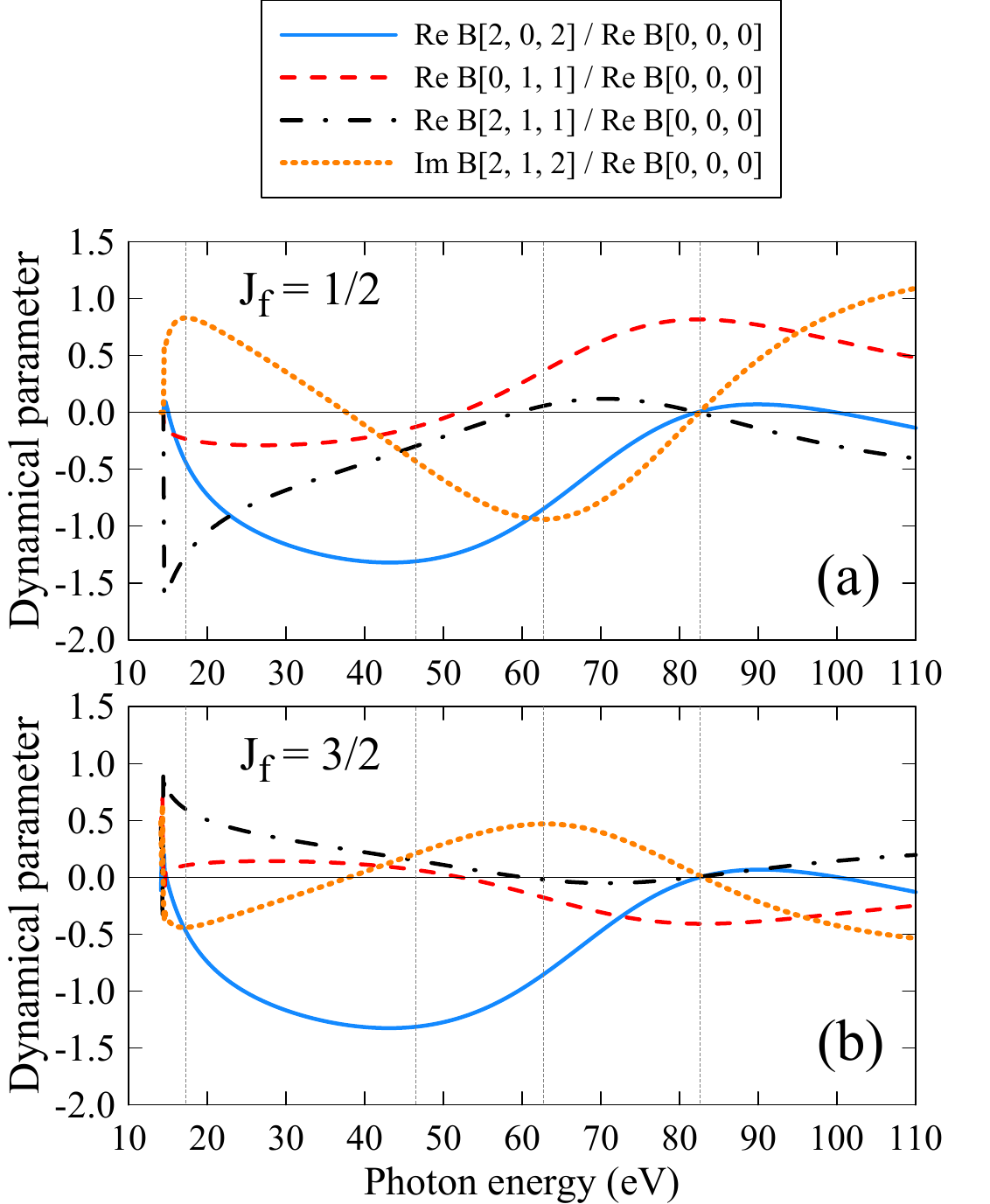}
\caption{Photon energy dependencies of the dynamical parameters relative to the value of $B[0,0,0]$ for final ionic states of Kr$^+\,4p^5$ with $J_f=1/2$ (a) and $J_f=3/2$ (b). Vertical dashed grey lines correspond to photon energies selected for analysis (see text).
\label{fig:Bpar}}
\end{figure}  

In Figure~\ref{fig:Bpar} we present the ratio of the dynamical parameters~(\ref{eq:Bparameter_general}) to zero-rank $B[0,0,0]$ because specifically these parameters determine the vector correlation characteristics. Quite typically that dipole angular anisotropy parameter $\beta\sim B[2,0,2]/B[0,0,0]$ behaves in the same way for the ionization into $J_f=3/2$ and $J_f=1/2$ final ionic state (blue line in Fig.~\ref{fig:Bpar}a,b), while spin polarization proportional to $B[k_l,1,k]/B[0,0,0]$ possesses an opposite sign and approximately twice smaller value for $J_f=3/2$ \cite{Fano1965}
. The component $B[k=\{0,2\},1,1]/B[0,0,0]$ is formed by {\it transfer of polarization} in terms of \cite{Cherepkov1983}, and appears due to the circularly polarized component of the radiation. The component $B[2,1,2]/B[0,0,0]$ is {\it {dynamical polarization}}, and appears due to the fine-structure splitting. It has been a subject of number of investigation that typically dynamical polarization is smaller than polarization transfer. The component $B[0,1,1]/B[0,0,0]$ (red dashed line in Fig.~\ref{fig:Bpar}) is the only one which conserves after integration over photo\-electron emission angle and therefore for the detectors which collect all of electrons. There is a Cooper minimum in the $\varepsilon d\,j$ ionization amplitude near the 80~eV, which governs the zero of $B[2,k_s,k]/B[0,0,0]$ (blue, orange, black lines in Fig.~\ref{fig:Bpar}a,b). In opposite, the component $B[0,1,1]/B[0,0,0]$ which contains $\varepsilon s\,1/2$ amplitude manifests a maximum at this energy. One can see that $B[2,0,2]/B[0,0,0]$ in Fig.~\ref{fig:Bpar}a approaches very close to the minimal possible value $-\sqrt{2}$ at energy 40~eV. To reach this value ratio of $\varepsilon s\,1/2$ and $\varepsilon d\,3/2$ amplitudes should be real and equal $1/\sqrt{2}$. Substituting this ratio into Eqs.~(\ref{eq:Bparameter_general}) one gets zero for all of the spin components. Therefore the crossing of all curves near the 40~eV is not occasional. For the $J_f=3/2$ (Fig.~\ref{fig:Bpar}b) there are two $\varepsilon d\, 3/2(5/2)$ amplitudes and the explanation is a bit trickier but still fair.

\begin{figure*}[!h]
\begin{center}
\includegraphics[width=17.5 cm]{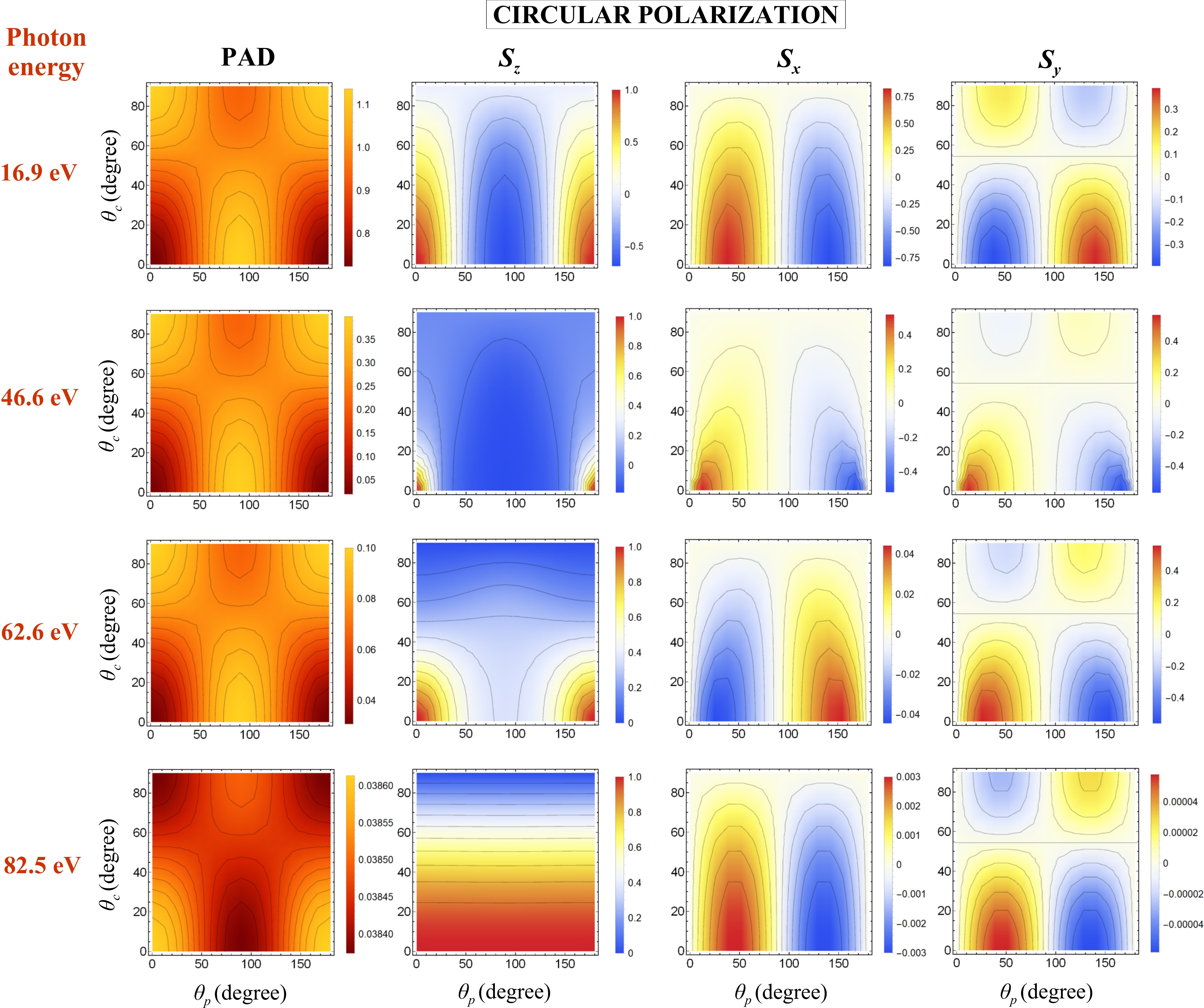}
\caption{PAD (first column) and photo\-electron spin component along the field propagation direction $S_z$ (second column), in the direction of electron momentum projection to the $xy$-plane $S_x$ (third column) and in the direction perpendicular to field propagation and electron momentum $S_y$ (fourth column) as a function of polar emission angle $\theta_p$ and cone angle $\theta_c$ in case of krypton atom ionization by circularly polarized Bessel light.
\label{fig:CIRC_results}}
\end{center}
\end{figure*}

In Figure~\ref{fig:CIRC_results} the PAD and spin components due to ionization of krypton by the circularly polarized Bessel light are presented as a function of the polar emission angle $\theta_p$ and cone angle $\theta_c$ for different photon energies marked in Figure \ref{fig:Bpar} by the vertical lines. Presented results are calculated for residual ion in the state $J_f=1/2$, the results for $J_f=3/2$ are easily deduced by comparison Figs.~\ref{fig:Bpar}a and~\ref{fig:Bpar}b. 

The probability of electron emission (Fig.~\ref{fig:CIRC_results}, first column) caused by plane-wave light ($\theta_c$ tends to zero) is maximal in the plane perpendicular to propagation direction, but the maximum turns to minimum when the cone angle increases. For the photon energy 82.5~eV the PAD is practically uniform because $\beta\approx 0$. The component parallel to the field propagation direction $S_z$ (second column) may form a quite complicate pattern if $B[2,1,1]/B[0,0,0]$ is essential: for this target ($4p$-shell of krypton) that corresponds to low energy 16.9~eV  (see Fig.~\ref{fig:CIRC_results}, upper row). When the ratio $B[2,1,1]/B[0,0,0]$ drops down the angular pattern became more uniform and  the minima at $\theta_p=90^{\circ}$ completely disappears at 82.5~eV. Thus the zero spin at $\theta_p=45^{\circ}$ for low energy (second column, first row) is dynamical and depends on the ionization amplitudes. Two other components lie on the plane perpendicular to the field propagation direction: $S_x$ (third column) is parallel to the projection of electron momentum to this plane, and $S_y$ (fourth column) is orthogonal. Both possess a maximum at $\theta_p=45^{\circ}$ according to equation~(\ref{eq:Circ_Sxy}). Remarkably that while $S_z$ and $S_x$ decreases with cone angle, the $S_y$ component has a more complicate pattern with  a zero point at $\cos{\theta_c}=1/\sqrt{3}$  (marked by the horizontal line) and  for the cone angle above this value the spin changes orientation. 

\begin{figure*}[!h]
\begin{center}
\includegraphics[width=17.5 cm]{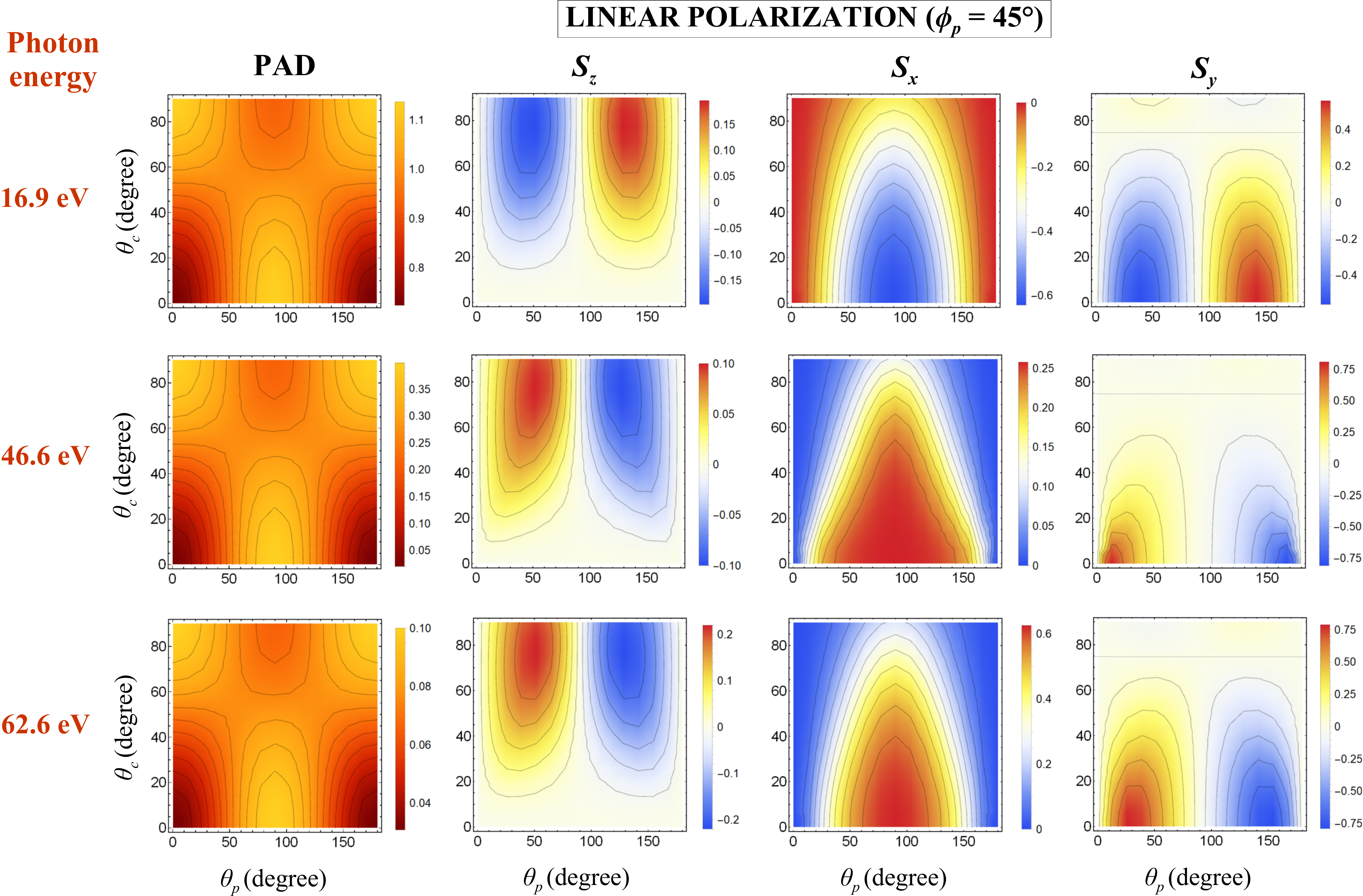}
\caption{PAD (first column) and photo\-electron spin component along the field polarization $S_z$ (second column), in the field propagation direction  $S_x$ (third column) and in the direction perpendicular to them $S_y$ (fourth column) as a function of polar emission angle $\theta_p$ and cone angle $\theta_c$ in case of krypton atom ionization by linearly polarized Bessel light. The  angle $\phi_p=45^{\circ}$ \label{fig:LIN_results_45dg}}
\end{center}
\end{figure*}

In Figure~\ref{fig:LIN_results_45dg} the PAD and spin components due to ionization of krypton by the linearly polarized Bessel light are presented as a function of polar emission angle $\theta_p$ and cone angle $\theta_c$ at the fixed azimuth angle $\phi_p=45^{\circ}$ for different photon energies marked in Figure~\ref{fig:Bpar} by the vertical lines (for residual ion in the state $J_f=1/2$). Since the photo\-emission caused by linearly polarized radiation, the transfer mechanism does not contribute and attitude of spin polarization is lower than in case of circularly polarized radiation. In ionization by the plane-wave linearly polarized radiation spin polarization can not be oriented neither along polarization vector nor along electron emission. The last component does not appear for the twisted radiation as well (see Eq.~(\ref{eq:Lin_spin})). Remarkably that spin orientation along polarization is allowed and increases with cone angle $\theta_c$ (see Figure \ref{fig:LIN_results_45dg}, second column). The other two component $S_{x,y}$ behave in different way as a function of the cone angle. Besides, $S_y$ manifests a zero at $\sin^2(\theta_c)=(4-\sqrt{2})/7$ and changes orientation after that. It is also interesting to note that PAD defined by eq.~(\ref{eq:PADstattens}) in case of linearly polarized Bessel beam coincides exactly with that of circularly polarized Bessel beam at $\phi_p=45^{\circ}$ (see first columns of Figs.~\ref{fig:CIRC_results} and~\ref{fig:LIN_results_45dg}). That directly (but not quite evidently at the same time) follows from the comparison of eqs.~(\ref{eq:Circ_PAD}) and (\ref{eq:Lin_PAD}).

\section{Conclusions}
\label{Section4}

In the current work we present an investigation of photo\-electron and photo\-ion polarization in an ionization caused by the twisted Bessel light. More precisely we obtain the equations for electron spin polarization, ionic orientation and alignment in terms of photo\-ionization amplitudes within $jj$-coupling scheme in general form and in the dipole approximation. The equations are applicable for the extended target with uniform distribution of atoms. The photo\-electron spin is considered in two coordinate systems related with photo\-electron emission and the beam propagation  direction. That allows us to analyse in the high details dependencies on the twisted light cone angle. 

As for plane-wave radiation in the case of circularly polarized beam there is spin component oriented along the beam propagation direction which conserves after integration over photo\-emission direction. In case of the twisted radiation, the angular dependency of this component dynamically connected with the cone angle. We show that lowest rank polarization parameter, e.g. ion orientation and the components complanar to the beam propagation and emission directions (conserved part of $S_z$ and tangential components) monotonically decreases with the cone angle, while the  orthogonal to the plane component behaves as the second Legendre polynomial, possesses a zero at $\cos\theta_c=1/\sqrt{3}$ and changes the orientation for the wider cone angle.

For the linearly polarized twisted radiation the results are even more intrigue: same as for plane-wave radiation the spin is orthogonal to the electron emission direction, but a component oriented along the polarization vector, which does not exist in the conventional plane-wave case, appears.


As an illustrative example we consider photo\-ionization of $4p$-shell of atomic krypton in the dipole approximation within $jj$-coupling scheme. It was found that even though the continuum is assumed to be smooth some interesting spin polarization components patterns as a functions of photo\-electron polar emission angle $\theta_p$ and Bessel beam cone angle $\theta_c$ are observed. The more specific energy (spectroscopic) dependency is observed for circularly polarized beam in component oriented along propagation direction. For the other component energy dependency follows for the value of the dynamical parameters.

\begin{acknowledgments}
The authors are acknowledge A.~Surzhykov for careful reading of the manuscript and useful suggestions. The work on the development of formalism for the photoelectron spin polarization in atomic ionization by twisted light and analysis in case of krypton atom were funded by the Russian Science Foundation (project No. 21-42-04412,~\cite{RSFproject}). The calculations of the photoionization amplitudes were performed using resources of the Shared Services “Data Center of the Far-Eastern Branch of the Russian Academy of Sciences” and supported by the Ministry of Science and Higher Education of the Russian Federation (project No. 0818-2020-0005) and Ministry of Science and Higher Education of the Russian Federation grant No. 075-15-2021-1353.
\end{acknowledgments}

\appendix


\bibliography{Twisted}

\end{document}